\documentclass[pre,preprint,showpacs,showkeys]{revtex4}

\begin{document}

%---------------------------------------------------------------
% Title & address
\title{Distribution of epicenters in the Olami-Feder-Christensen
model}

\author{Tiago P. Peixoto}
\email{tpeixoto@if.usp.br}
\author{Carmen P. C. Prado}
\email{prado@if.usp.br}
\affiliation{Instituto de F\'{\i}sica, Universidade de S\~ao Paulo \\
Caixa Postal 66318, 05315-970 - S\~ao Paulo - S\~ao Paulo - Brazil}

\date{\today}

% ----------------------------------------------------------------
% Abstract

\begin{abstract}

We found that the  well established Olami-Feder-Christensen (OFC) model for
the dynamics of earthquakes is able to reproduce a new striking
property of real earthquake
data.  Recently, Abe and Suzuki found that the
epicenters of earthquakes could be connected to generate  a graph,
with properties of a  scale-free
network of the Barab\'asi-Albert type. However, only the non
conservative version of the 
Olami-Feder-Christensen model is able to reproduce this behavior. The
conservative version, instead, behaves like a random graph. Those
findings, besides indicating the robustness of the model to describe
earthquake dynamics, reinforce that conservative and nonconservative
versions of the OFC model are qualitatively different, 
and  propose a completely new dynamical
mechanism that, without an explicit rule of preferential attachment,
is able to generate a free scale network. The preferential
attachment is in this case a ``by-product''  of the long term correlations built by the
self-organized critical state. We believe that the detailed study of
the properties of this network can reveal new aspects of the dynamics
of the OFC model, contributing to the understanding of self-organized
criticality in non conserving models.

\end{abstract}

%-----------------------------------------------------------------
% Pacs e keywords


\maketitle

%--------------------------------------------------------------------
% Corpo do artigo

The concept of self-organized criticality was introduced by Bak, Tang
and Wiesenfeld \cite{Bak} in 1987, as a
possible explanation of scale invariance in nature. To illustrate
their basic ideas, they presented a cellular
automaton model, the sandpile model, so called  because of a possible
analogy between its dynamical rules and the movement of sand or snow in
avalanches. Since this seminal work, a great number of
cellular automata and coupled map models have been investigated, in an
attempt to elucidate the essential
mechanisms hidden in such a wide class of different non-linear
phenomena whose statistics of events (or
avalanches) are governed by  power-laws. However, up to now, one still lacks from a
general theoretical framework for self-organized criticality. Success
in  analytical investigations have been achieved in many models. For a
revision see, for instance, \cite{book,turcotte}.

In this context, a model that has been widely studied in the literature is the
Olami-Feder-Christensen (OFC) model for the dynamics of earthquakes. The original
OFC model, introduced in 1992 \cite{OFC},  is a two-dimensional coupled 
map model defined on a square lattice,
whose dynamical rules were inspired in a spring-block model proposed to describe the dynamics of
earthquakes. Earthquakes, in the real world, are associated with many 
power-laws, the most known
of them being the Gutenberg-Ritcher law for the distribution of
avalanche energies.
 The OFC model assigns  - to each site of a square lattice - a real
variable $z_{i,j}$ (energy or tension), initially chosen at random in
the interval $[0,z_c)$, where $z_c$ is a threshold value. 
$z_{i,j}$ increases
slowly throughout the lattice and each time that, for a given site, 
$z_{i,j}$ exceeds  $z_c$,  the system relaxes. A fraction $\alpha \,
z_{i,j}$ of the
tension of site $(i,j)$ is then distributed to each of its nearest
neighbors. As a consequence, the tension of some of its neighbors
may also exceed $z_c$, generating an `avalanche' that will only stop
when $z_{i,j}<z_c$ again for all sites of the lattice. We have assumed
 open boundaries in our simulations.

Within the OFC model there is a dissipation parameter
$\alpha$. If $\alpha=0.25$ the total tension in the lattice, $\sum
z_{i,j}$, is conserved during the
avalanching process, in the bulk of the lattice (there is always 
dissipation in the boundaries). But if $\alpha < 0.25$ there
is some dissipation also in the bulk of the system. Because of those 
facts, this model has been widely studied in
literature: it is, at the same time, a prototype of self-organization 
in systems with non-conservative relaxation
rules (the existence of SOC in the non-conservative models is, up to
now,  not well
understood  \cite{Prado,Chris,Lise,Miller}) and also a paradigm of the success of SOC ideas, since
it is able to reproduce important aspects of the dynamics of 
earthquakes. 

Recently, Abe and Suzuki \cite{Abe1} observed a new power-law in
the statistics of earthquakes. They analyzed earthquake data from
both the district of southern California and Japan,
connecting their epicenters in order to generate a graph. Each area
analyzed was divided into small cubic cells; they  associated to each
of these cells a node  every time an earthquake started inside it. 
The epicenters of two
successive earthquakes were linked,  defining  an edge. 
In this way the data has been mapped into a complex growing graph that  behaves like a scale-free
network of the Barab\'asi-Albert type \cite{Barabasi}. The degree distribution
of the graph decays as a power-law. The clustering coefficient
and the diameter of a cluster were also calculated, showing
small-world network properties \cite{Watts}. These features have revealed a novel aspect of
earthquakes as a complex critical phenomenon. 

In this paper we studied the Olami-Feder-Christensen model to see if it could also predict
this new striking behavior. We found that the non conserving version of the
model  reproduces the behavior of experimental data,
even for a very small degree of non conservation.
The degree distribution  of the evolving network formed by
its epicenters  is scale free. 
However, the conservative version of the model  has a
qualitatively different behavior, more similar  to a random graph, whose
degree distribution is Poisson, indicating that most of the nodes have
the same degree and - although random - the corresponding network  is much more
homogeneous. These results are in agreement with some recent
observations, reinforcing that conservative and non conservative
versions of the OFC model are quite different.
Hergarten and Neugebauer (2002) \cite{Hergarten} studying
the efficiency of the OFC model to predict foreshocks and aftershocks,
de Carvalho and Prado (2003)\cite{Josue}, studying the transient behavior of the
OFC model and  Miller and Boulter (2003) \cite{Miller03} studying the
distribution of values at which supercritical sites topple
have also reported  qualitatively different behaviors between  the
conservative and non conservative OFC model. 

In a complex graph, the edges are not distributed in an regular way and
not all nodes have the same number of edges. 
One possible way to characterize complex
networks is through its  distribution function $P(k)$, which gives the
probability that a random selected node has exactly $k$ edges. $k$ is
called the {\it degree} of the node. In a
random graph, since the edges are placed randomly among the nodes, the
majority of nodes have approximately the same degree, close to the
average connectivity $\langle k \rangle$, and the 
distribution $P(k)$ is a Poisson
distribution with a peak at $P(\langle k \rangle)$. 
Most complex networks, however,
have a distribution function $P(k)$ that deviates  significantly from
a Poisson distribution. In particular, for a large number of networks,
associated with a wide class of systems, ranging from the world wide web
to metabolic networks, $P(k)$ has a power-law tail, $P(k) \sim
k^{-\gamma}$. Such networks are called scale free \cite{Barabasi}, and
have called the attention of many researchers in the last years.

We simulated the OFC model in a square lattice,  building  graphs with a procedure very
similar to what has been employed by Abe and Suzuki. Each site that
gives birth to a new avalanche is an epicenter; each epicenter 
defines a node, and every node is then connected to the node where 
the next epicenter occurs, establishing a link or edge between
them. After many avalanches this procedure
generates a complex network (or graph), and we have studied some of 
its statistical properties.

After eliminating a transient of at least $10^6$ events,  we calculate numerically the 
distribution function $P(k)$ for the graph constructed from the time
sequence of epicenters in the OFC model,  for different values of $\alpha$ and
different lattice sizes. As the first and last sites are the only ones
with an odd number of edges, they were eliminated. Our results for the
distribution $P(k)$ can be seen 
in figure 1. It is clear that,  if $\alpha <
0.25$ (figure 1a), the distribution is scale-free for some
decades, with an
exponent $\gamma$ that varies linearly with $\alpha$ (see figure 2), at
least for values of $\alpha$ not too far from the conservative regime.  
The network grows toward the inside of the
lattice, with the most connected sites in the borders and the most
inner sites being the last ones being added to it (see figure 3a). 
The complex structure, however, is not a boundary effect. If we take out
the border sites and adjusts the scale, we see that the same spatial
structure is reproduced (figure 3b). 
Because  one needs  a growing 
network to observe the scale free-behavior \cite{Barabasi}, after a
certain number of events, as a consequence of
the finite size of the lattice, most of the sites of the lattice  
have already become part of the network. At this point the
scale free behavior starts to break.

If the system is conservative, however, the distribution function
$P(k)$ has a well defined
peak, indicating a higher degree of homogeneity among the nodes 
(figure 1b). Figures 4a and 4b, that shall be compared with 
figure 3, shows the spatial distribution of connectivities in the
lattice. As expected, it is much more homogeneous. This homogeneous
behavior is not destroyed if we vary the statistic of events.

Finally, our findings seems also to be robust with respect to the cell
size. If we increase the size of the cell, defining, for instance,  four  adjacent
sites of the lattice as a unique cell, there is no change in the results, not even in
the exponent $\gamma$ that characterizes the  degree distribution $P(k)$, as shown in
figure 5.

In conclusion, we have shown that the non conservative version of the
Olami-Feder-Christensen model is able to reproduce the scale free
network associated to the dynamics of the epicenters observed 
on real earthquake data.  The conservative version of the model
displays a qualitatively different behavior, being more close to a random graph. The
smallest degree of non conservation seems to be enough 
to change the behavior of the model, since  for $\alpha=0.249$ we see
that  $P(k)$ has already a well defined power law behavior for some 
decades. Those
findings, besides giving an indication of the robustness of this model
to reproduce the dynamics of earthquakes, reproducing the experimental
findings of Abe and Suzuki, present a completely new dynamical
mechanism  to generate a free scale network. There is no explicit rule of
preferential attachment, and the preferential attachment
observed in the network is a signature  of the model dynamics. We hope
that a complete study of the properties of the network can help to
solve some still controversial aspects of the Olami-Feder-Christensen
model and of self-organized critical behavior, and can be interesting
and useful
even if a more detailed study of earthquake data comes to show in the
future that
the the results reported by Abe and Suzuki are not universal.

%------------------------------------------------------------------------
% Bibliografia

%
%
%--------------------------------------------------------------
% figure captions
\newpage

\begin{center}
\bf{ Figure Captions}
\end{center}

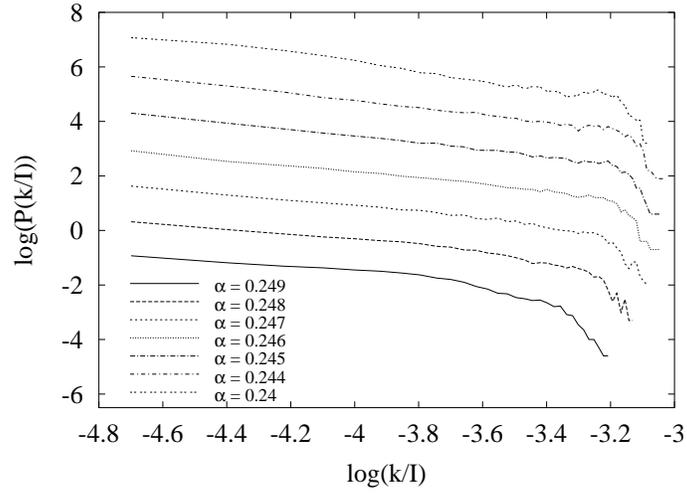
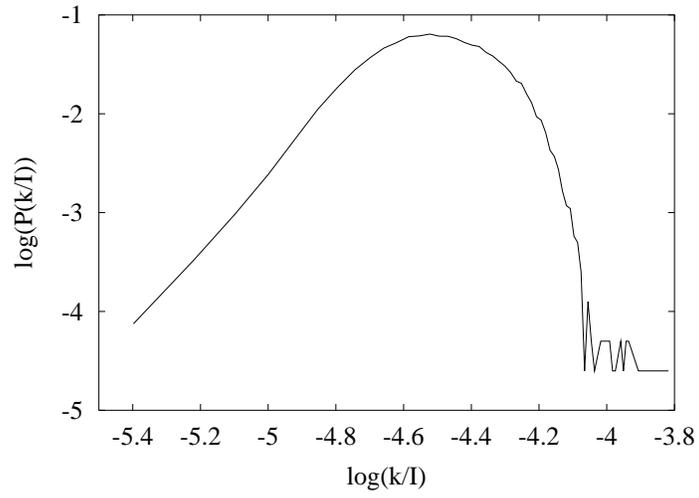
{\bf Figure 1:}  Normalized degree distribution $P(k/I)$ for
different values of $\alpha$. $I$ is the total number of epecenters. 
(a) non conservative regime: the results
show a free scale network behavior in all cases.
The curves for $\alpha < 0.249$ have been shifted upwards along the $y$
axis for clarity, otherwise they would all coincide. In all
cases  $L=200$ and the number of epicenters registered is $10^5$. (b) 
Conservative
regime: the degree distribution is similar to a random graph; in this 
case we have $L=200$
and an statistics of $10^6$ events. Lowering the statistics does not
change this behavior.

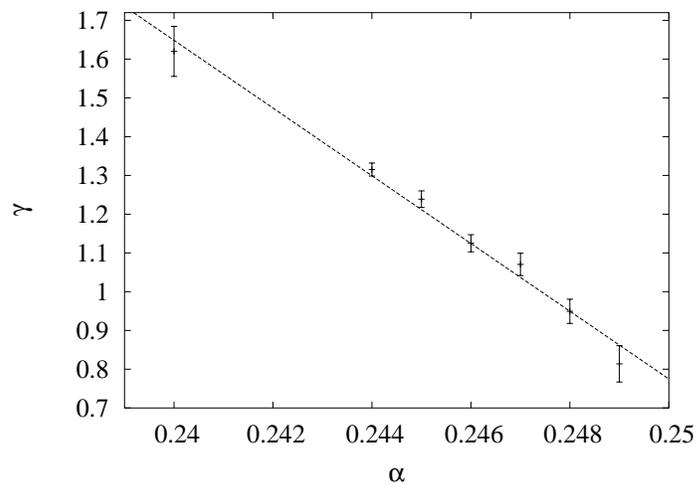
{\bf Figure 2:} Exponent $\gamma$, that characterizes the power law
behavior of $P(k)$, for different values of $\alpha$. $\gamma$ seems
to increase linearly with $\alpha$. In
all cases  $L=200$ the number of epicenters is  $10^5$.

{\bf Figure 3:} Spatial distribution of node degrees in the non
conservative case, for $\alpha=0.249$, $L=200$ and $10^5$ events. Sites
associated with nodes of higher degree are darker and, as one can
see, are
closer to the boundaries.   Figure (b) is a blow up of (a).  The $20$
sites closer to the boundaries have not been not plotted and the scale
has been 
changed in order to show the details. We can see that the
structure of the network is reproduced and is not a boundary effect. 

{\bf Figure 4:} Spatial distribution of node degrees for the
conservative case.  Sites associated with nodes of higher 
degree are darker, $L=200$ and the number of epicenters is $10^6$. 
(a) The same scale of figure 3a has been used. (b) The scale has
been
changed to reveal details of the structure of the network that, in this
case,  is much more homogeneous and quite different than the one
observed in the non conservative regime. 

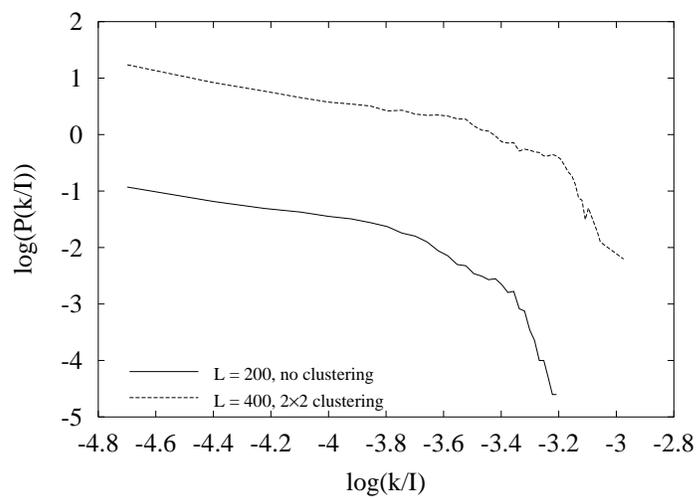
{\bf Figure 5:} The normalized degree distribution $P(k/I)$ for $\alpha=0.249$,
$L=200$ and $10^5$ events,  for different cell sizes. $I$ is the total
number of epecenters. (a) $L=200$ and
each site of the lattice defines a cell. (b) $L=400$ and each four
adjacent sites are in the same cell. The curve has been shifted upwards
in the axis $y$ for clarity.

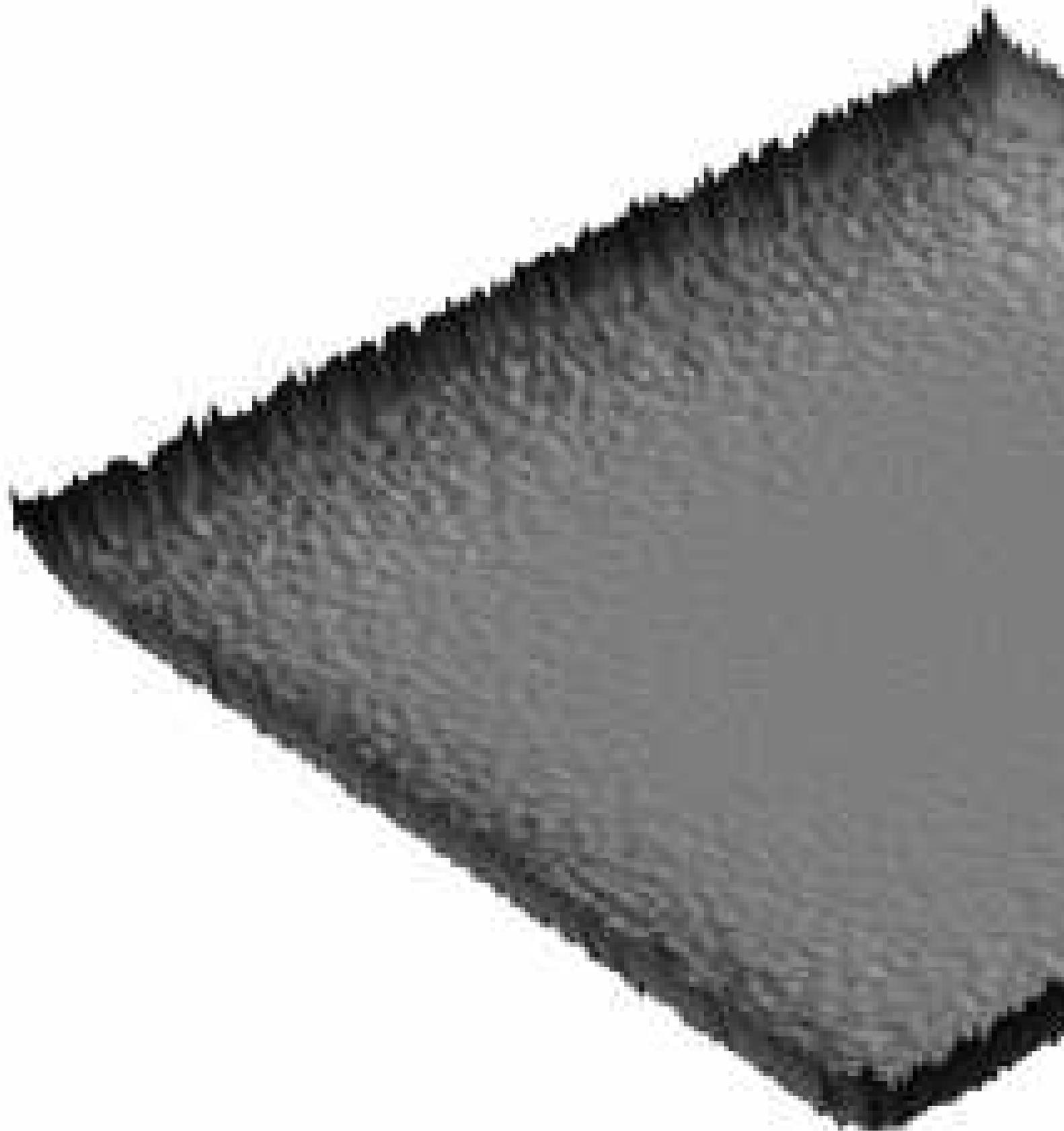
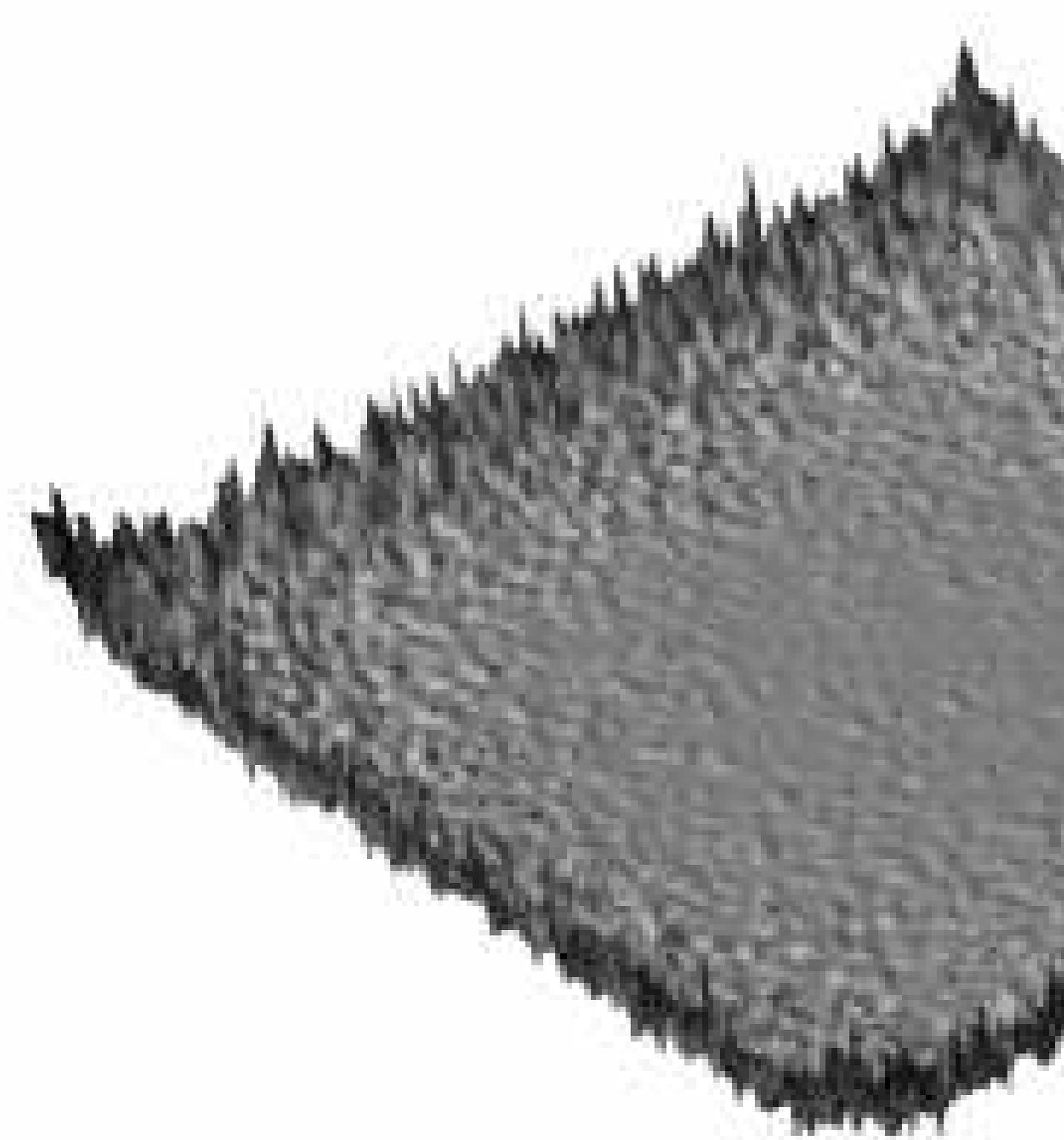
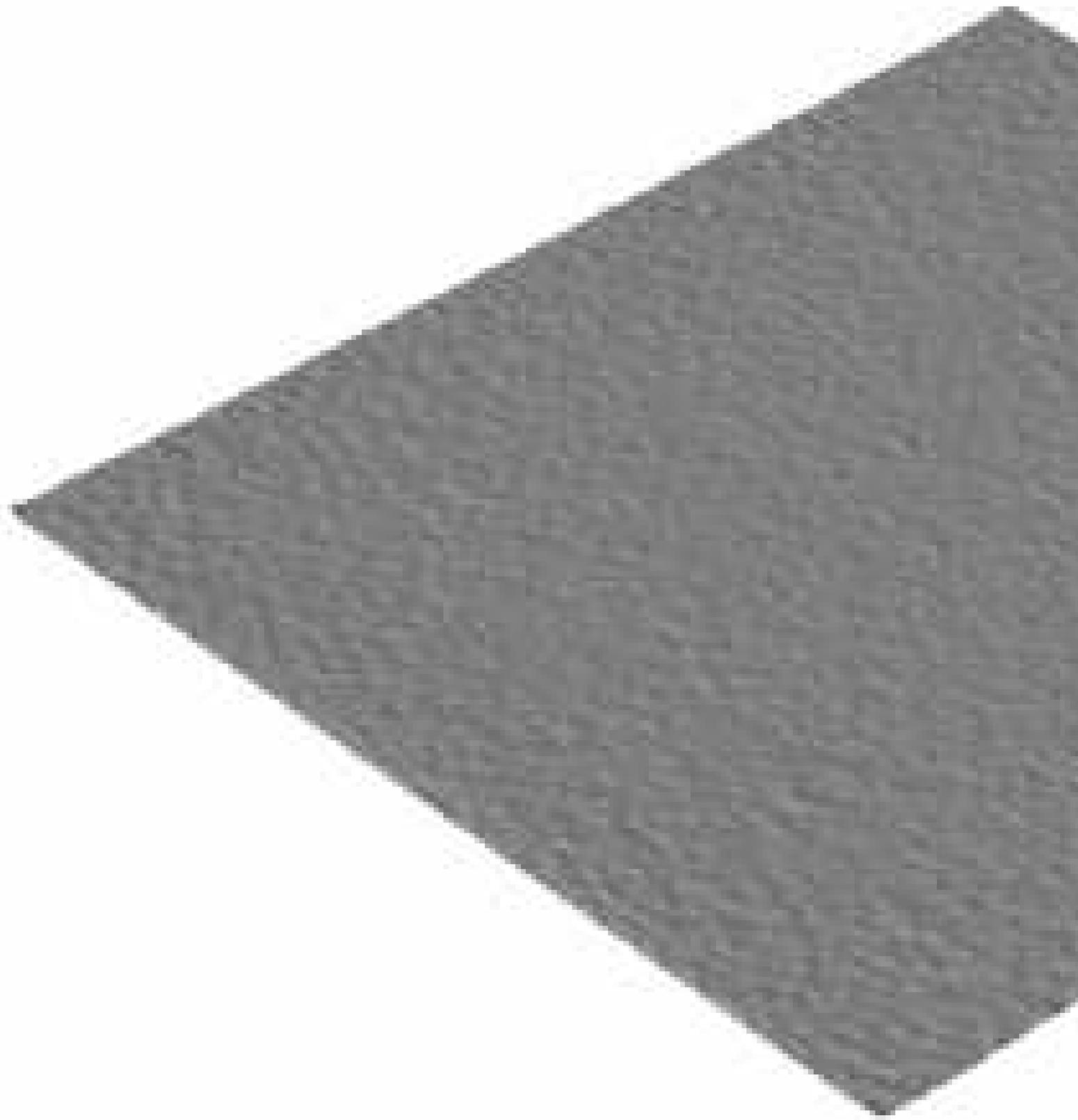
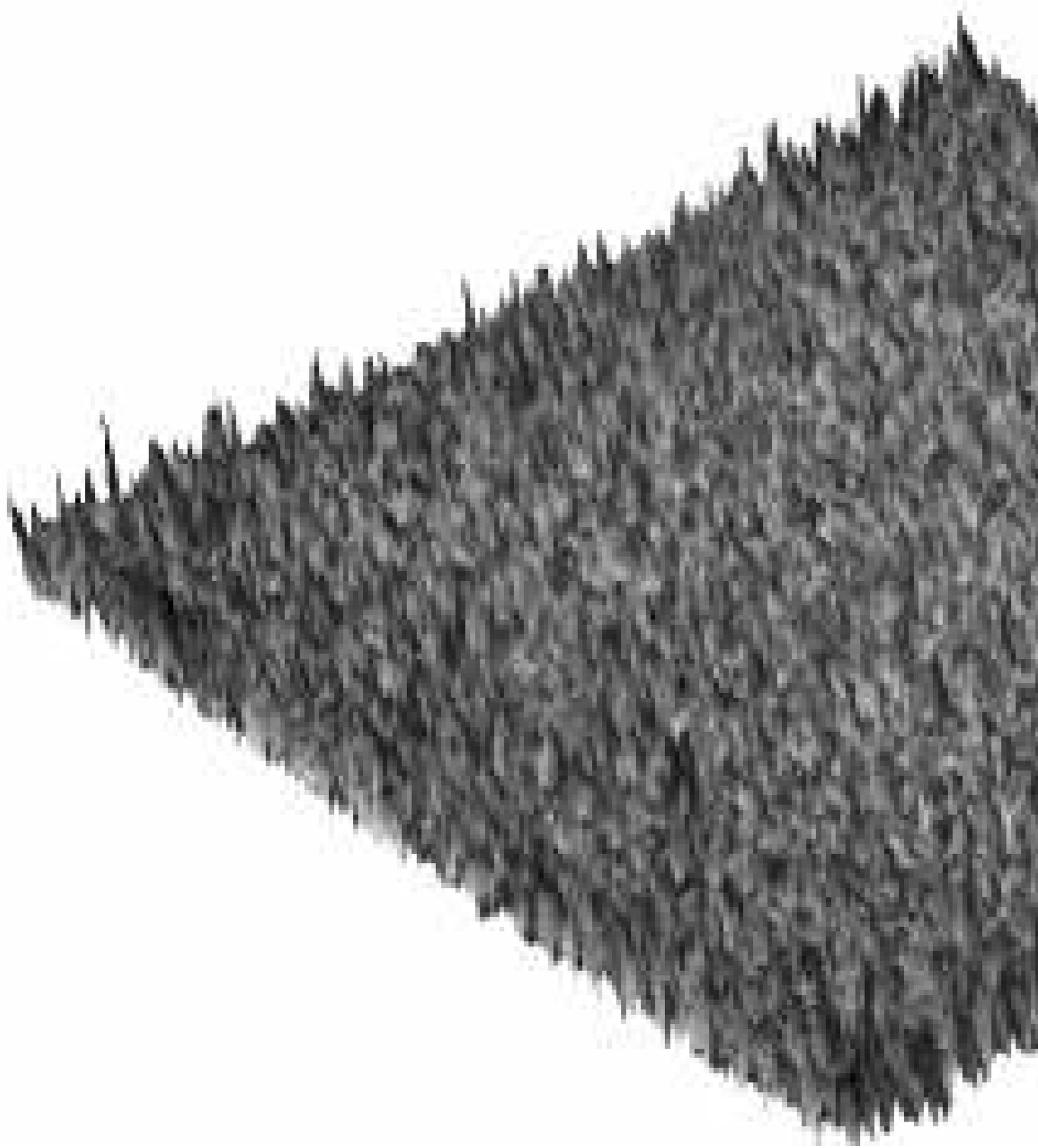
\end{document}